\documentclass[a4paper,twocolumn,nofootinbib,superscriptaddress]{revtex4-1}
\usepackage{hyperref}
\usepackage[T1]{fontenc}
\usepackage[utf8]{inputenc}
\usepackage{lmodern}
\usepackage{amsmath}
\usepackage{amsfonts}
\usepackage{amsthm}
\usepackage{bbold}
\usepackage[retainorgcmds]{IEEEtrantools}
\usepackage{graphicx}
\usepackage{tikz}
\usepackage[justification=RaggedRight]{caption}
\usepackage{subfigure}
\usepackage{color}
\usepackage{booktabs}
\usepackage{multirow}
\usepackage{pifont}

\newcommand{\tick}{\ding{52}}
\newcommand{\xmark}{\ding{55}}

\definecolor{nred}{rgb}{0.7,0.2,0.2}
\definecolor{nblack}{rgb}{0,0,0}
\definecolor{nblue}{rgb}{0.2,0.2,0.8}
\definecolor{ngreen}{rgb}{0.2,0.6,0.2}

\usepackage{paralist}

\begin{document}

\theoremstyle{definition}
\newtheorem{mydef}{Definition}
\newtheorem{mylemma}{Lemma}

\renewcommand{\today}{\number\day\space\ifcase\month\or
   January\or February\or March\or April\or May\or June\or
   July\or August\or September\or October\or November\or December\fi
   \space\number\year}

\begin{center}
\title{A tripartite quantum state violating the hidden influence constraints}

\date{\today}

\author{{Tomer Jack Barnea}}
\email[e-mail: ]{tomer.barnea@unige.ch}
\affiliation{Group of Applied Physics, University of Geneva, CH-1211 Geneva 4, Switzerland}
\author{Jean-Daniel Bancal}
\affiliation{Group of Applied Physics, University of Geneva, CH-1211 Geneva 4, Switzerland}
\affiliation{Centre for Quantum Technologies, National University of Singapore, 3 Science drive 2, Singapore 117543}
\author{Yeong-Cherng Liang}
\affiliation{Group of Applied Physics, University of Geneva, CH-1211 Geneva 4, Switzerland}
\author{Nicolas Gisin}
\affiliation{Group of Applied Physics, University of Geneva, CH-1211 Geneva 4, Switzerland}

\begin{abstract}

The possibility to explain quantum correlations via (possibly) unknown causal influences propagating gradually and continuously at a finite speed $v > c$ has attracted some attention recently. In particular, it could be shown that this assumption leads to correlations that can be exploited for superluminal communication. This was achieved studying the set of possible correlations that are allowed within such a model and comparing them to correlations produced by local measurements on a four-party entangled quantum state. Here, we report on a quantum state that allows for the same conclusion involving only three parties.

\end{abstract}

\maketitle

\end{center}

\section{Introduction}
\label{1}

Despite a lot of research, the question as to how non-local, i.e. Bell inequality violating~\cite{Bell} quantum correlations arise in space-time is still poorly understood.

Ever since the fundamental work of Bell~\cite{Bell} we know that equipping the particles with local (hidden) variables~\cite{bell66,EPR} and hence shared randomness or local common causes will not suffice. Experimental evidence demonstrating the existence of such non-local correlations, even when the measurements are space-like separated, has long been provided (see e.g. Ref.~\cite{Aspect99} and references therein), reawakening our concern about Einstein's famous sentence on the "spooky action at a distance"~\cite{Einstein}.

If we want to cling to the hope of providing a local and continuous causal explanation of such correlations in space and time, we have to consider explanations that go beyond adding only shared randomness. Different such attempts were proposed, i.e. in Ref.~\cite{Eberhard} by Eberhard or in Refs.~\cite{Scarani02,Scarani05} by Scarani and Gisin, where it was proposed that superluminal yet finite-speed influences carrying information about measurements performed could account for these correlations. These influences are presumably hidden, i.e. unknown to present day physics and their speed is defined with respect to some privileged reference frame. Moreover, the physical carrier of such influences should obey the principle of continuity, i.e. propagate gradually and continuously through space and time, which leads to the finite-speed assumption. However, these papers demonstrated that in such a case and if no additional local variables are involved, the influence could not remain hidden and could be exploited for superluminal communication.

Very recently the proof could be extended to include the case where additional local variables are allowed too~\cite{bancal}. For this purpose the notion of $v$-causal models was introduced therein (see also Refs.~\cite{gisin,rudolph,scarani13}). In such models correlations between measurement outcomes -- such as those leading to quantum violations of Bell inequalities~\cite{Bell} -- are expected to be a direct consequence of common causes and causal (superluminal) influences traveling at a finite speed $v$.

There are different ways to test $v$-causal models. One possibility is to conduct experiments with two highly synchronized parties and see if a Bell-inequality violation can be observed. This, however, has not ruled out the possibility of such kinds of models as it is limited by the synchronization precision possible in a laboratory. It has therefore only yielded lower bounds on the speed $v$, as was done in Refs.~\cite{Scarani00,salart,cocciaro,pan}. Nevertheless those hypothetical privileged reference frames moving at a constant speed with respect to Earth were tested in Refs.~\cite{salart,cocciaro,pan} using an idea by Eberhard~\cite{Eberhard}.

A different approach was chosen in Refs.~\cite{Scarani02,Scarani05} and pursued in Ref.~\cite{bancal}. In the latter it was shown that the non-local character of quantum theory in combination with $v$-causal models leads to (superluminal) signaling, i.e. the spatially separated observers in a Bell-type experiment can communicate with the arbitrarily distant parties by simply manipulating their measurement choices and observing their local outcome statistics. In other words, these influences trying to explain quantum non-locality \textit{cannot} remain hidden, in the sense that they can be exploited by the observer for superluminal communication. This allows us to disprove such models under the plausible assumption of a world, where faster-than-light communication is not possible. The argumentation was established with a quantum state that involved four parties each equipped with two measurement settings that produce dichotomic outputs.

No tripartite quantum state could be found at that stage and it remained an open question as to whether there is a fundamental difference between the three and the four-party cases. Note however that a tripartite solution with supra-quantum correlations was already known as was shown in Ref.~\cite{coretti}. However, the result in Ref.~\cite{coretti} cannot be verified using quantum physics, not even with ideal measurements. It requires supra-quantum correlations that, according to today's physics, do not exist. Hence our result is a major advancement.

In addition, the possibility to test the inequality upon which our result depends experimentally is an important issue. To be able to underline this result it would be very favorable to find a quantum state whose preparation is technically less involved and that is more robust to noise in order to conduct a real experiment. The combination of the evidence provided by such an experiment and the work mentioned above would exclude a lot of directly conceivable explanations of quantum correlations.

This work reports on a tripartite quantum state found that allows for the same conclusion as the four-party example in~\cite{bancal}. In Sec.~\ref{2} we formally introduce $v$-causal models. A tripartite Bell-like inequality that has to be satisfied by non-signaling $v$-causal models and its quantum violation is presented in Sec.~\ref{3}. Finally we give a conclusion in Sec.~\ref{4}.

\section{$v$-causal models}
\label{2}

\begin{figure}[htb]
\begin{center}
\includegraphics[width = 0.7\columnwidth]{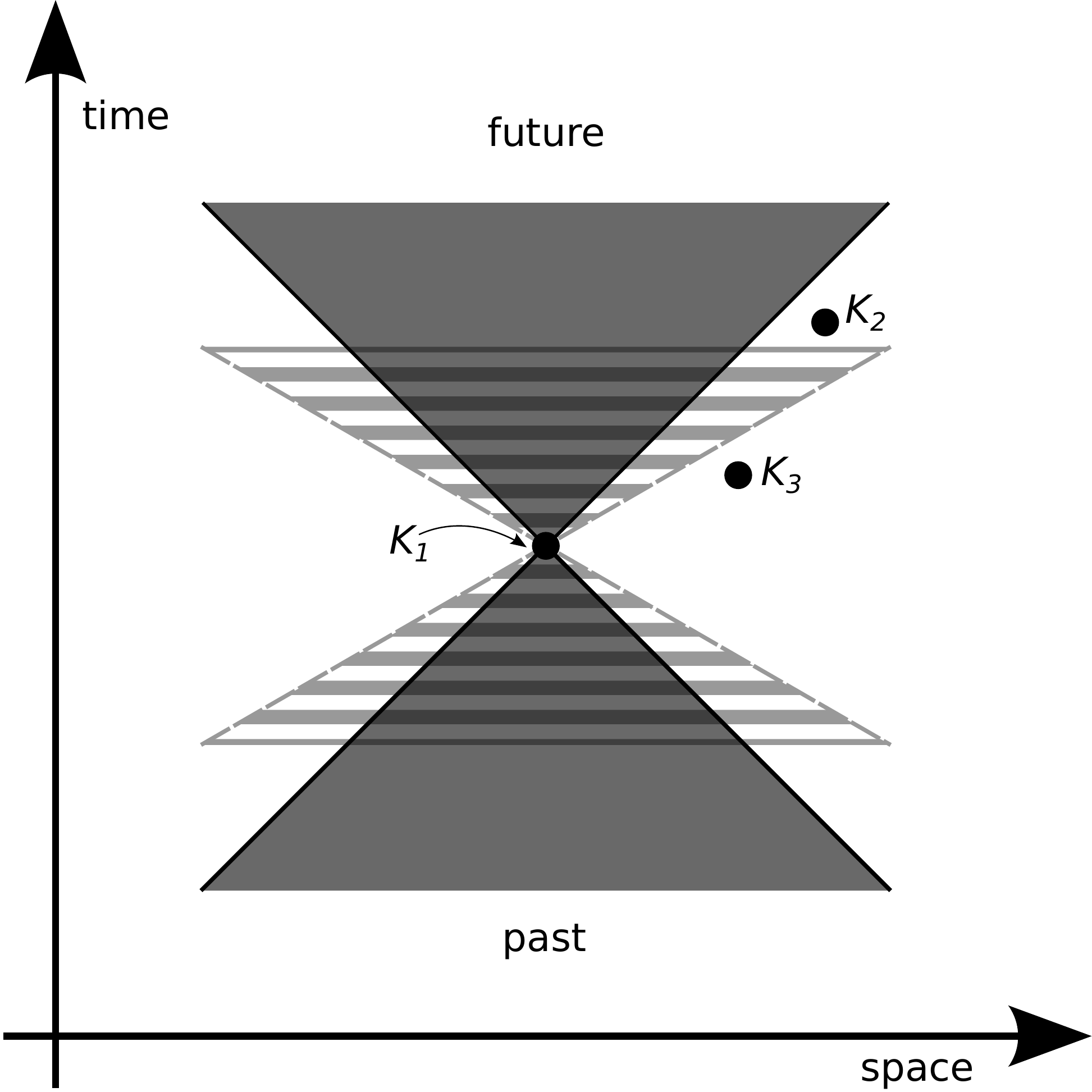}
\caption{Space-time diagram in the preferred reference frame illustrating the concepts of $v$-connected and $v$-disconnected events. The shaded region are the $c$-cones (light cones) and the striped regions show the $v$-cones with $v > c$. The $v$-cones of $K_{1}$ contain all points that can reach $K_{1}$, or that can be reached from $K_{1}$, by signals (influences) propagating at a speed not larger than $v$.}
\label{vcone}
\end{center}
\end{figure}

We now recall the key ingredients of a $v$-causal model, as introduced in Ref.~\cite{bancal}. A $v$-causal model seeks to provide explanations of correlations between measurement events, such that they obey the principle of continuity. In other words from the measurement location information spreads gradually and continuously through space and time. Additionally such a model is equipped with shared randomness. We thus first have to establish the different time orderings between these events. Bear in mind that in a $v$-causal model influences can propagate at a speed $v > c$, presumably in some preferred -- yet unknown -- reference frame\footnote{See Refs.~\cite{bohm,cocciaro12} and references therein for a discussion of such a reference frame.}. We thus cannot rely on the causal structure of relativity. Rather, causality is defined in the preferred frame, as represented in the space-time diagram of Figure~\ref{vcone}.

Here, an important distinction has to be made between the relation of $K_{1}$ and $K_{2}$ on the one hand and $K_{1}$ and $K_{3}$ on the other. $K_{2}$ lies in the future $v$-cone of $K_{1}$ and can therefore be causally influenced by $K_{1}$ via influences propagating at speed $v$. These two events are \textit{$v$-connected}. We write this as $K_{1} < K_{2}$. On the contrary, $K_{3}$ lies neither in the future nor in the past $v$-cone of $K_{1}$. These events are styled \textit{$v$-disconnected} and denoted by $K_{1} \sim K_{3}$. Formally, a $v$-causal model for two measurement events $A$ and $C$ is one that satisfies the following requirements.

\begin{mydef}
\label{vcausal}
A \textbf{$v$-causal model} for the measurement events $A$, labeled by the measurement setting (input) $x$ and the outcome (output) $a$ and 
$C$ with input $z$ and outcome $c$ is one that satisfies the equations below
\begin{IEEEeqnarray}{l?l}
P_{A<C}(ac|xz,\xi) = & \sum_{\lambda} q(\lambda|\xi) P(a|x\lambda) P(c|z,ax\lambda) \IEEEyesnumber\IEEEyessubnumber\\ \label{vcausala}
P_{C<A}(ac|xz,\xi) = & \sum_{\lambda} q(\lambda|\xi) P(c|z\lambda) P(a|x,cz\lambda) \IEEEyesnumber\IEEEyessubnumber\\ \label{vcausalb} 
P_{A \sim C}(ac|xz,\xi) = & \sum_{\lambda} q(\lambda|\xi) P(a|x\lambda) P(c|z\lambda) \IEEEyesnumber\IEEEyessubnumber \label{vcausalc}
\end{IEEEeqnarray}
\end{mydef}

Here $\lambda$ can be understood as a complete characterization of a region in the intersection of the past $v$-cones of $A$ and $C$ that suffices to make predictions about them and $q(\lambda)$ is its probability distribution~\cite{bellcuisine,norsen}, cf. Figure~\ref{bac}. As opposed to the usual locality condition we included an additional parameter in this definition to account for the fact that these decompositions can be conditioned on additional relevant information in the intersection of the past $v$-cones of $A$ and $C$ (see Appendix C of Ref.~\cite{bancal} for a more detailed discussion in the four-party scenario). In particular, in the presence of an auxiliary party -- as we will see in Section~\ref{3} -- $\xi$ can take the value of this party's input and/or output.

Note that in the first two cases a $v$-causal model is capable to reproduce arbitrary quantum correlations. Indeed, we make the working hypothesis that quantum correlations observed so far are produced by $v$-connected measurement events in the preferred frame. On the other hand, a $v$-causal model can only produce local, i.e. non-Bell-inequality-violating correlations in the case where $A \sim C$. This follows from the assumption that in a $v$-causal model Bell-inequality violation arises from these causal influences propagating at finite speed $v$. In the aforementioned case these influences will not arrive on time and therefore we cannot observe non-local correlations.

Clearly, we would like to have a $v$-causal model that reproduces Bell-inequality-violating quantum correlations, as that was the goal to begin with. As explained above this is not possible in cases where $A \sim C$, because our model predicts local correlations. If we restrict ourselves to $v$-connected events it is possible to attain a consistent definition of a $v$-causal model for quantum correlations, see Appendix A in Ref.~\cite{bancal}.

\section{The tripartite case}
\label{3}

\subsection{Preliminaries}

\begin{figure}[!]
\subfigure[The unconditioned time ordering]{\label{acb}\includegraphics[width=0.6\columnwidth]{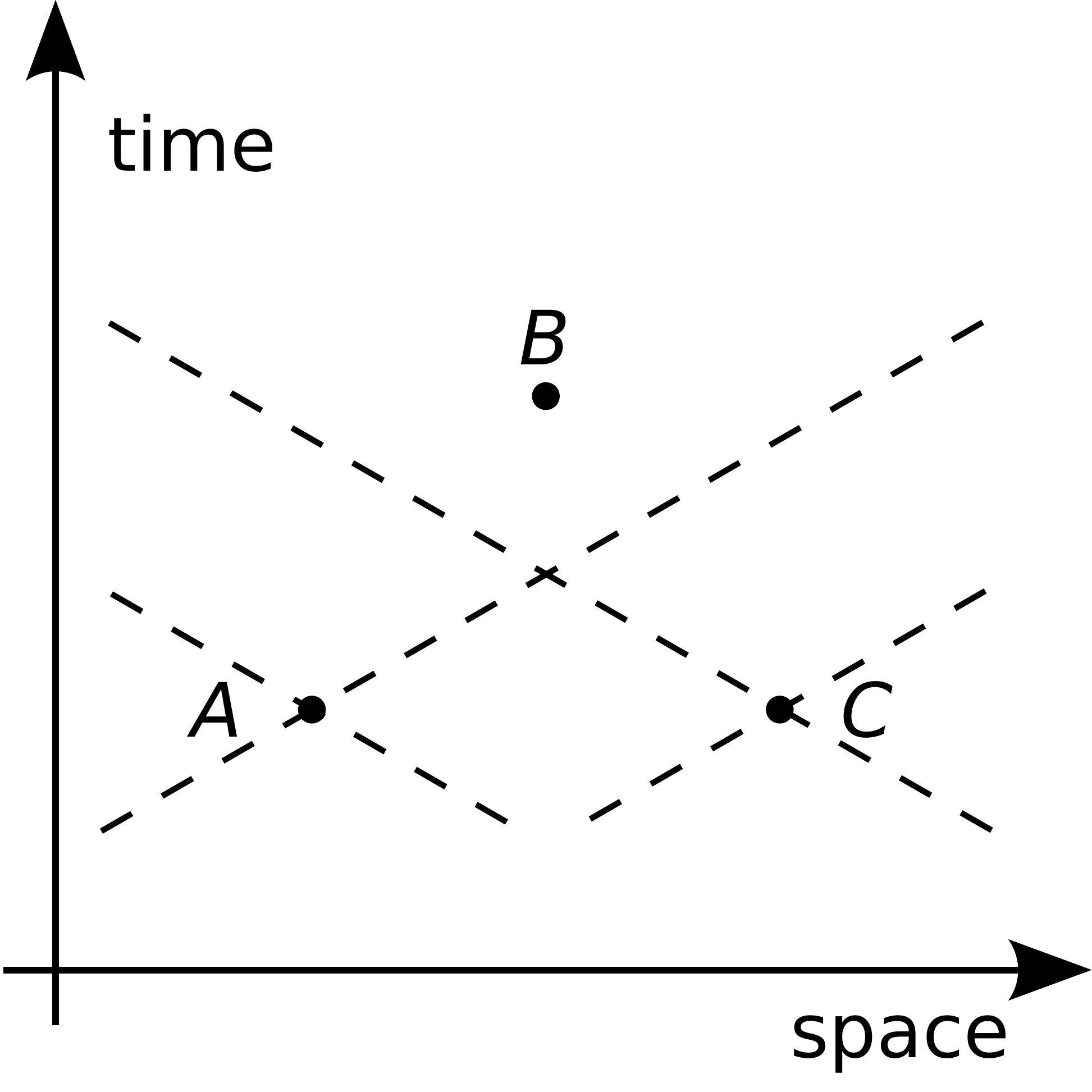}}
\subfigure[The conditioned time ordering]{\label{bac}\includegraphics[width=0.6\columnwidth]{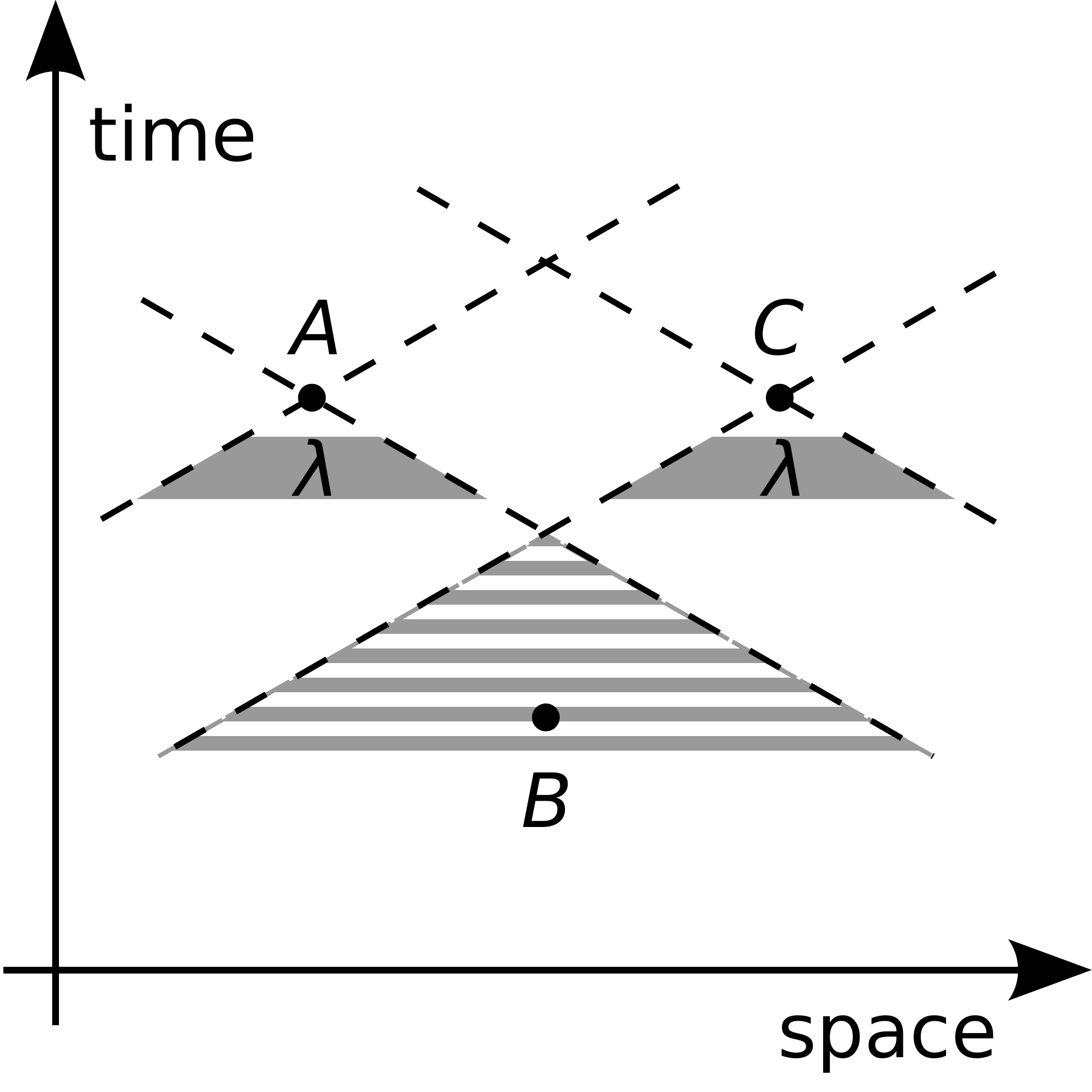}}
\caption{Space-time diagrams showing the two relevant time orderings for the tripartite scenario. Here, $A$ and $C$ are simultaneous in the hypothetical preferred frame of reference. In (b), the common past $v$-cones of $A$ and $C$ also include $B$, labeled by input $y$ and output $b$.}
\label{scen}
\end{figure}

To fully understand the next section we briefly recall the important steps followed in Ref.~\cite{bancal}. The space-time configuration is very important for the argumentation. Up to permutations of parties there are qualitatively two different cases to consider\footnote{In reality there are four different temporal configurations of measurement events. However in the case where all parties are $v$-connected $v$-causality imposes no constraint, and when all parties are mutually $v$-disconnected no multipartite quantity need to be quantum. These configurations cannot be used to demonstrate the signaling property of $v$-causal models in the way we do it here.}: The two $v$-disconnected parties are either measured before or after the third party that is $v$-connected to both of them (as shown in Figure~\ref{scen}). The major difference between these two cases manifests itself in the way to impose the locality constraints for the disconnected parties cf. Eq.~\eqref{vcausalc}. If the two $v$-disconnected parties measure before the single party, see Figure~\ref{acb}, the locality constraints between them cannot be conditioned on later events, e.g. the measurement of the single party. In other words $\xi$ cannot contain information about the measurement settings and outcomes of the single party, and we might equally take this variable out of the picture by choosing $\xi=\emptyset$. In the other case, see Figure~\ref{bac}, what happens to the sole party $B$ is contained in both parties' past $v$-cones and thus the locality constraints can be conditioned on his input $y$ and output $b$: we can choose $\xi=(b,y)$.

We will refer to these scenarios as \textit{unconditioned} and \textit{conditioned}, respectively. Since the unconditioned locality condition can always be decomposed as a sum of conditioned ones, it is more general to consider the conditioned case. The analysis as well as the results presented in this paper were carried out considering the conditioned scenario, we will therefore restrict ourselves to this case from now on.

The next task consists of characterizing the set of correlations that can be described within a $v$-causal model for the chosen space-time configuration and which do not allow for superluminal communication. A sufficient condition for some tripartite correlations $P(abc|xyz)$ not to be exploitable for superluminal communication through the manipulation of inputs is that they satisfy a series of mathematical constraints known as the non-signaling conditions~\cite{pr,barrett05}. Essentially, in the tripartite case, these conditions stipulate that the bipartite marginal correlations are well-defined and are independent of the input of the remaining party, i.e.
  \begin{IEEEeqnarray}{C}
  \label{ns}
  \sum\limits_{c} P(abc|xyz) =P(ab|xy) \quad \forall\quad a,b,x,y,z 
  \end{IEEEeqnarray}
and similarly for the other parties. Moreover, as is evident from the specified space-time ordering given in Figure~\ref{bac}, no non-local correlations between the parties $A$ and $C$ can be achieved as they lie outside each others $v$-cones. The desired set of non-signaling correlations described by $v$-causal models is thus a convex polytope~\cite{gruenbaum} defined by Eq.~\eqref{ns} and all the locality constraints imposed between parties $A$ and $C$, cf Eq.~\eqref{vcausalc}\footnote{Formally, the locality constraints that have to be imposed in the space-time ordering of Figure~\ref{scen}(b) are captured by lifted Bell inequalities~\cite{Pironio}.}.

To compare the correlations allowed by non-signaling $v$-causal models against those of quantum theory, a second step consists of restricting our attention to observable quantities that do not involve simultaneous measurements. We want to avoid simultaneity, because -- as mentioned above -- we do not expect quantum correlations to be reproduced by $v$-causal models in this case. Technically, this amounts to projecting away variables involving simultaneous measurements, e.g. the $AC$ correlation terms in the case where $A\sim C$. The set of probability distributions obtained in this way will henceforth be referred to as the \textit{projected hidden influence polytope}. Using this terminology the task can be rephrased in the following way. Characterize the projected hidden influence polytope for three parties and find a quantum probability distribution outside of this set.

\subsection{A tripartite hidden influence inequality and its quantum violation}

We considered the scenario where the two $v$-disconnected parties $A$ and $C$ have two inputs and outputs while the other party $B$ has three of each.
For such a setting the following lemma is valid.

\setdefaultenum{(i)}{(a)}{i.}{A.}
\setdefaultleftmargin{3em}{}{}{}{}{}

\begin{mylemma}
\label{lem}
For a tripartite probability distribution $P(abc|xyz)$ with $a$,$c$,$x$,$z$ $\in \{0,1\}$ and $b$,$y$ $\in \{0,1,2\}$ that
\begin{compactenum}
  \item satisfies the non-signaling constraints~\eqref{ns}, \\
  \item is local
  between the parties $A$ and $C$ conditioned on $B$, i.e.
  \begin{IEEEeqnarray}{C}
  \label{loc}
  P(ac|xz,by) = \sum\limits_{\lambda} q(\lambda|by) P(a|x\lambda) P(c|z\lambda)
  \end{IEEEeqnarray}  
\end{compactenum}
the following inequality $S$ holds:
\end{mylemma}

\begin{IEEEeqnarray}{rCl}
\label{S}
S & = & -\: P_{A}(0|0) - P_{A}(0|1) - P_{B}(0|0) \nonumber\\
&& +\: P_{B}(1|0) - P_{B}(0|1) - P_{B}(1|1) \nonumber\\
&& +\: P_{AB}(00|00) + P_{AB}(00|10) + P_{AB}(01|00) \nonumber\\
&& -\: P_{AB}(01|10) + 2P_{AB}(00|01) + 2P_{AB}(01|01) \nonumber\\
&& +\: 2P_{AB}(00|12) + P_{AB}(01|12) + P_{C}(0|0) \nonumber\\ 
&& +\: P_{BC}(00|00) - P_{BC}(00|10) - P_{BC}(10|10) \nonumber\\
&& -\: P_{BC}(00|20) - P_{BC}(10|20) - P_{BC}(10|01) \nonumber\\
&& +\: P_{BC}(00|11) + P_{BC}(10|11) - P_{BC}(00|21) \nonumber\\
&& \geq -2
\end{IEEEeqnarray}

\textbf{Proof} Define $P_{B}(2|0) = \sum_{a,c} P(a2c|x0z) $ to be the marginal probability $P(b=2|y=0)$ and let $P_{AC|B}(ac|xz)$ denote the $AC|B$ marginal probabilities $ P(ac|xz,b=2,y=0)$, similarly for $P_{A|B}$ and $P_{C|B}$. That inequality~\eqref{S} holds can be seen by rewriting the above expression as follows\footnote{This can be done using the non-signaling conditions, Eq. \eqref{ns}, the normalization of probabilities and the definition of marginal probabilities.}:

\begin{IEEEeqnarray}{rCl}
\label{I}
I & = & P_{ABC}(000|101) + P_{ABC}(000|011) \nonumber\\
&& +\: P_{ABC}(010|011) + P_{ABC}(000|100) \nonumber\\
&& +\: P_{ABC}(000|000) + P_{ABC}(010|000) \nonumber\\
&& +\: P_{ABC}(110|121) + P_{ABC}(120|121) \nonumber\\
&& +\: P_{ABC}(120|010) + P_{ABC}(120|120) \nonumber\\
&& +\: P_{ABC}(001|120) + P_{ABC}(011|120) \nonumber\\
&& +\: P_{ABC}(001|121) + P_{ABC}(001|010) \nonumber\\
&& +\: P_{ABC}(011|010) + P_{ABC}(001|001) \nonumber\\
&& +\: P_{ABC}(011|001) + P_{ABC}(121|011) \nonumber\\
&& +\: P_{ABC}(111|100) + P_{ABC}(111|101) \nonumber\\
&& +\: P_{B}(2|0) [1-P_{A|B}(0|1)-P_{C|B}(0|0) \nonumber\\
&& \quad +\: P_{AC|B}(00|00) + P_{AC|B}(00|10) \nonumber\\
&& \quad -\: P_{AC|B}(00|01) + P_{AC|B}(00|11)] \nonumber\\
& = & S + 2
\end{IEEEeqnarray}

The first 20 terms in Eq.~\eqref{I} and the one in front of the square brackets are probabilities and therefore non-negative and the expression in the square brackets is exactly the Clauser-Horne (CH) expression~\cite{ch} for the parties $A$ and $C$ conditioned on $B$ and therefore by condition $(ii)$ this term is non-negative too. In summary we can conclude that $I \geq 0$ and therefore $S \geq -2$ if conditions $(i)$ and $(ii)$ of Lemma~\ref{lem} are satisfied. \hfill $\Box$ \\

Let us return to the two crucial steps to establish our argument. By looking at the expression for the inequality $S$ we notice that it does not involve any term with both the parties $A$ and $C$. Taking into consideration the locality condition in the lemma we have indeed found an inequality for the projected hidden influence polytope for the space-time configuration of Figure~\ref{bac} as well as Figure~\ref{superluminal}.

The next task comprises of finding a quantum state violating the inequality $S$ shown above. It turns out that a quantum state $|\Psi \rangle_{ABC} \in \mathbb{C}^2 \otimes \mathbb{C}^3 \otimes \mathbb{C}^2 $ suffices. Indeed considering the quantum state given in Eq.~\eqref{aa} and the measurement in Eq.~\eqref{bb} one finds a small but clear\footnote{Although the value might seem small, it lies well above the numerical precision. This can be verified using the explicit state and the measurements given in Appendix~\ref{A}.} violation of inequality~\eqref{S} that amounts to -2.00015.\footnote{From the converging hierarchy~\cite{navascues07a,navascues08a,doherty08a} of semidefinite programs, one can check that the strongest possible quantum violation of inequality~\eqref{S} is of the same order of magnitude, namely -2.0003. This violation can be achieved using a similar state and settings to those given in Appendix~\ref{A}, but using more significant digits.}

Let us briefly explain why the combination of these two steps indeed allows us to see that $v$-causal models for quantum theory are signaling, i.e. do not obey Eq.~\eqref{ns}. First of all note that -- as mentioned in the paragraph above -- there exists a quantum violation for inequality~\eqref{S}. Secondly, a $v$-causal model can reproduce the quantum marginals $P(ab|xy)$ and $P(bc|yz)$ in the configuration considered (of Figure~\ref{bac}), because if one of $A$ or $C$ decided to postpone his measurement, the $BC$ or $AB$ marginal would remain unchanged and would have to be the correlations issued from $|\Psi \rangle_{ABC}$ since all parties would be $v$-connected. However inequality~\eqref{S} only involves those marginals for a violation, it must thus be violated by the $v$-causal model as well, and hence either one of the conditions in Lemma~\ref{lem} must be wrong. Given the fact that condition $(ii)$ holds for a $v$-causal model per definition (recall that this is always the case if $A$ $\sim$ $C$), we must conclude that condition $(i)$ has to be dropped, making the model signaling which was the claim.

\subsection{From signaling to superluminal communication}

Having established a violation of the non-signaling constraints let us try to gain a deeper insight of the exact meaning behind this statement. First and foremost the probability distributions we are dealing with will not fulfill the relations given in Eq.~\eqref{ns}. In this formal definition of non-signaling the speed of light never appears. The justified question arises as to how one can achieve superluminal communication from a violation of these constraints. If we can choose the space-time configuration of the involved parties freely this is always possible. Indeed, a violation of the non-signaling constraints implies that the outputs of certain parties depend on the input of the other party. So by placing the single party at a large enough distance from the other parties, the first party can transmit information by merely changing his input.

\begin{figure}[!]
\includegraphics[width=0.6\columnwidth]{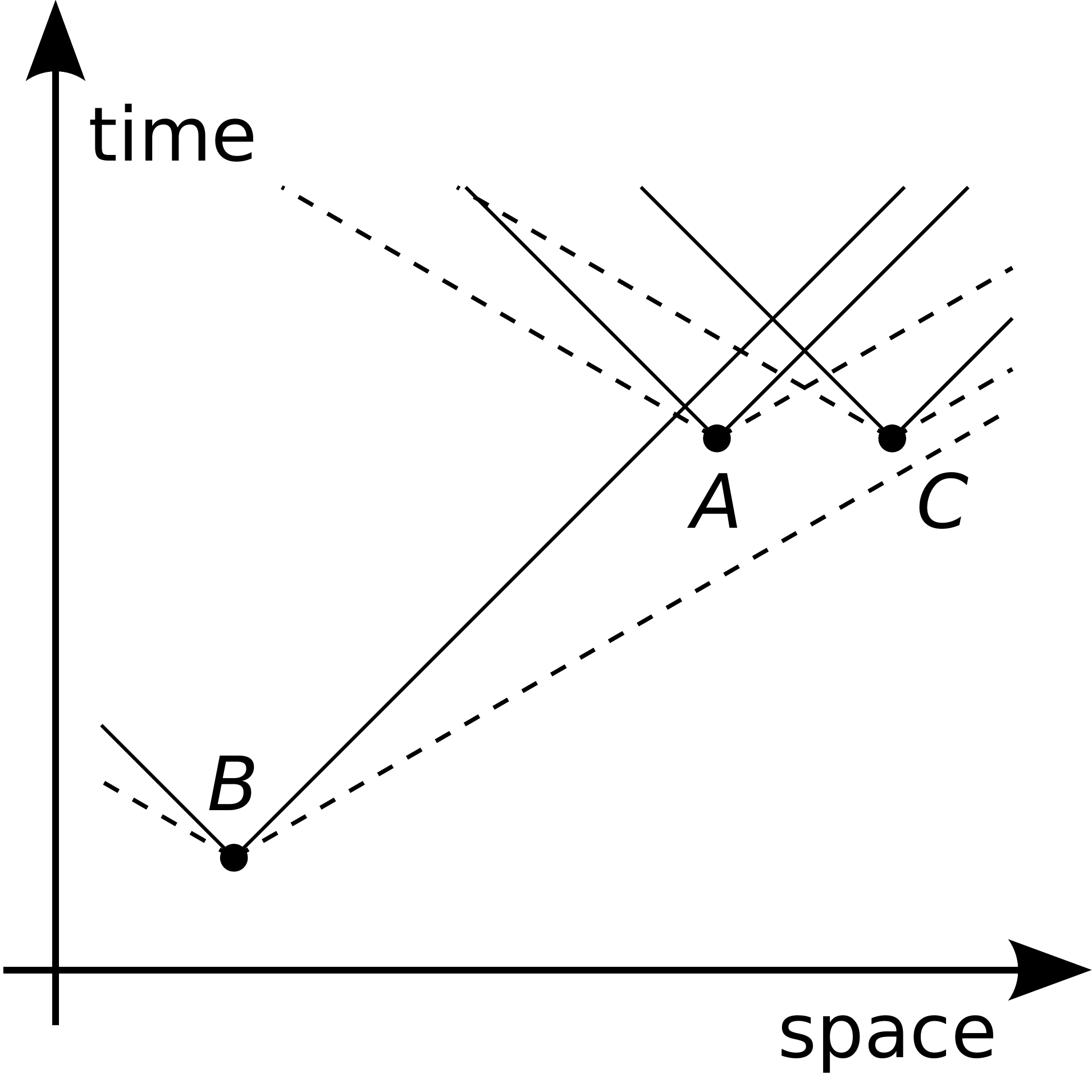}
\caption{The space time configuration to demonstrate superluminal communication for correlations $P(ac|xz,by)$ that depend on $y$ and $b$. The solid lines represent the $c$-cones and the dashed lines the $v$-cones, respectively.}
\label{superluminal}
\end{figure}

However let us recall the space-time configuration discussed in Figure~\ref{bac}. In the tripartite case there are three different possibilities for signaling to surface. Either the $AB$ marginal depends on $C$’s input or the $AC$ marginal depends on $B$’s input or the $BC$ marginal depends on $A$’s input. In our case only the second solution is possible because both the $AB$ and the $BC$ marginals are -- by assumption -- quantum and therefore non-signaling.

Hence, in the situation of Figure~\ref{bac}, there is no hope to achieve superluminal signaling by varying the inputs of $B$ since there is no point outside the future light-cone of $B$ at which the $AC$ marginal can be evaluated. But we can create such a point if we change the space-time configuration, positioning the parties as shown in Figure~\ref{superluminal}, thus using the dependence of the $AC$ marginal on $B$ to carry information outside of $B$'s future light cone. Note also that the reasoning we followed to demonstrate a violation of the non-signaling condition~\eqref{ns} in the configuration of Figure~\ref{bac} can be reproduced to show that this condition is also violated in this new configuration.

Note also that in Figure~\ref{superluminal}, $A$ and $C$ can be chosen as close to each other, and as close to the border of Bob's $v$-cone as desired. In this way, superluminal communication at a speed arbitrarily close to that of the hidden influences $v$ can be obtained.

\subsection{Related and intermediate results}

Independently of the case we just presented in which all parties use binary inputs and outputs, except for Bob who has three of each, we also considered simpler Bell scenarios. In the first and most obvious one where all parties have dichotomic inputs and outputs, we could show in the conditioned case that the projection of the non-signaling and the hidden influence polytope coincide. The unconditioned scenario can be ruled out by appealing to the monogamy of non-signaling correlations~\cite{toner,masanes} as was already pointed out in Ref.~\cite{coretti}. Here, however, we could rule out the possibility of quantum or non-signaling violations in the conditioned case as well. As quantum states are non-signaling there is no hope to find a state violating the hidden influence constraints if these two polytopes are identical.

\begin{table}[hbtp]
\begin{tabular}{lcc}
\toprule
\multirow{2}{*}{Scenario} & \multicolumn{2}{c}{Violation} \\
& N-S & Quantum \\
\midrule
\{[2 2] [2 2] [2 2]\} & \xmark & \xmark \\
\{[3 3] [3 3] [3 3]\} & \tick & \bf{?} \\
\{[3 2] [2 2] [2 2]\} & \xmark & \xmark \\
\{[2 2] [3 2] [2 2]\} & \tick & \bf{?} \\
\{[2 2] [3 3] [2 2]\} & \tick & \bf{?} \\
\{[2 2] [3 2] [3 2]\} & \tick & \bf{?} \\
\{[2 2] [3 3] [3 2]\} & \tick & \bf{?} \\
\{[2 2] [3 3 3] [2 2]\} & \tick & \tick \\
\bottomrule
\end{tabular}
\caption{Summary of the results of a selection of studied cases. The Bell-type scenarios are labeled as follows: The number of square brackets corresponds to the number of parties and the number of entries therein is the number of inputs for that party. The actual value in the square bracket stands for the number of outputs for that particular input. In the first and the third scenario listed above, the corresponding projected hidden influence polytope can be solved completely and no violation from quantum nor correlations respecting Eq.~\eqref{ns} (abbreviated as N-S) is possible. For all the other scenarios shown, it is possible to find violations with N-S correlations, but a quantum violation (which also implies a N-S violation) was found only in the last case.
}
\label{tab1}
\end{table}

Without any results in the easiest case a broadening to either more inputs or outputs was necessary. Equipping all parties with three outputs for each of their two inputs we could establish a difference between the projected hidden influence polytope and the non-signaling one. This difference also emerges if instead of increasing the number of outcomes, the number of inputs is set to three~\cite{pironiopriv}; see also~\cite{coretti} for an example with four inputs. Studying the case with more outcomes in more depth we could observe that neither all parties nor all inputs necessarily utilized the maximum number of three outputs at their disposal. Hence the features of some intermediate cases were explored as well. A summary of the results achieved can be found in Table~\ref{tab1}.

\section{Conclusion}
\label{4}

The first reported tripartite quantum state that forces any $v$-causal model for quantum correlations to be signaling was described. As mentioned in previous sections Bell tests involving only two parties have only set a lower bound for the speed of such hypothetical influences. Therefore this result closes the gap between what has been experimentally achieved in the two-party case and what has been theoretically demonstrated in the four-party scenario.

Albeit with a difference to the four-party case where two inputs and outputs per party were enough to conclude the argument, this was not achievable in the tripartite case. Nevertheless, this finding shows that there is no fundamental difference between three and four parties in what concerns refuting $v$-causal models.

We also note that the argumentation involved for our tripartite example, as well as the results presented in Ref.~\cite{bancal}, do not rely on the "transitivity of non-locality" (as formulated in Ref.~\cite{coretti}). Indeed, the marginal correlations $AB$ that are involved in our example can be easily shown to satisfy all Bell inequalities.

From an experimental perspective, this work has to be seen as a proof of principle, as the weak violation and hence the low robustness to noise of the reported quantum state makes an experimental test exceedingly demanding. It remains an open question as to whether a quantum state violating the hidden influence constraints can be found that is robust against noise as well as easily producible experimentally.

\begin{acknowledgments}

The authors thank V. Scarani for comments on an earlier version of this paper. This work was supported by the Swiss State Secretariat for Education and Research through the COST Action MP1006, the European ERC AG Qore, the CHIST-ERA DIQIP and the Swiss NCCR QSIT as well as the National Research Foundation and the Ministry of Education, Singapore.

\end{acknowledgments}

\bibliography{/home/tomy/Documents/Figures/mycollection}

\appendix

\section{The tripartite quantum state}        
\label{A}
Inequality~\eqref{S} can be violated with the quantum state and the measurements given below. The superscript denotes the different inputs, while the subscript gives information about the outputs. Only the positive-operator-valued measure (POVM) elements for all but the last output are given; the remaining one can be obtained from the fact that POVM elements for the same input must sum to the identity operator. Since some coefficients are given numerically, we use the notation $\propto$ to mean that a state should be multiplied by a coefficient in order to be normalized.

\begin{IEEEeqnarray}{rCr}
\label{aa}
  |\Psi\rangle \propto & & (-0.0003-0.0075i)|001\rangle \nonumber\\
  & + & (0.0029+0.0093i)|010\rangle \nonumber\\
  & + & (0.6769)|021\rangle \nonumber\\
  & + & (-0.1145-0.2881i)|100\rangle \nonumber\\
  & + & (-0.5782-0.3330i)|111\rangle \nonumber\\
  & + & (-0.0154+0.0055i)|120\rangle
\end{IEEEeqnarray}

\begin{IEEEeqnarray}{rCl}
\label{bb}
        \hat{A_{0}^{0}} & = & \frac{1}{2}(\mathbb{1} + \sigma_{z}) \nonumber \\
        \hat{A_{0}^{1}} & = & P_{\alpha_0^1},\ \alpha_0^1 \propto {\begin{bmatrix} 0.5289 - 0.4693i \\ 0.7071 \end{bmatrix}} \nonumber \\
        \hat{B_{0}^{0}} & = & P_{\varphi_{0}^{0}},\ \varphi_{0}^{0} \propto {\begin{bmatrix}  0.0368 - 0.2164i \\ 0.7070 \\ 0.6584 + 0.1357i \end{bmatrix}} \nonumber \\	 
        \hat{B_{1}^{0}} & = & P_{\varphi_{1}^{0}},\ \varphi_{1}^{0} \propto {\begin{bmatrix}  -0.0368 + 0.2164i \\ 0.7072 \\ -0.6583 - 0.1357i \end{bmatrix}} \nonumber \\
        \hat{B_{0}^{1}} & = & P_{\varphi_{0}^{1}},\ \varphi_{0}^{1} \propto {\begin{bmatrix}  0.1466 - 0.0131i \\ 0.9891 \\ -0.0001 - 0.0027i \end{bmatrix}}  \nonumber \\	 
        \hat{B_{1}^{1}} & = & P_{\varphi_{1}^{1}},\ \varphi_{1}^{1} \propto {\begin{bmatrix}  0.9889 \\ -0.1467 - 0.0131i \\ 0.0006 - 0.0181i \end{bmatrix}}  \\
        \hat{B_{0}^{2}} & = & P_{\varphi_{0}^{2}},\ \varphi_{0}^{2}  \propto {\begin{bmatrix}  0.0002 + 0.0053i \\ 0.6925 - 0.1428i \\ 0.7072 \end{bmatrix}}  \nonumber \\	 
        \hat{B_{1}^{2}} & = & P_{\varphi_{1}^{2}},\ \varphi_{1}^{2}  \propto {\begin{bmatrix}  1.0000 \\ 0 \\ -0.0003 + 0.0075i \end{bmatrix}}  \nonumber \\
        \hat{C_{0}^{0}} & = & P_{\gamma_0^0},\ \gamma_0^0 \propto {\begin{bmatrix} 0.4800 - 0.8751i \\ 0.0619 \end{bmatrix}} \nonumber \\
        \hat{C_{0}^{1}} & = & P_{\gamma_0^1},\ \gamma_0^1 \propto {\begin{bmatrix} 0.0298 - 0.0543i \\ 0.9981 \end{bmatrix}} \nonumber
\end{IEEEeqnarray} 
	
\end{document}